\documentclass[page-classic]{epl2} %for one column style
\newcommand{\be}{\begin{equation}}
\newcommand{\ee}{\end{equation}}
\newcommand{\bea}{\begin{eqnarray}}
\newcommand{\eea}{\end{eqnarray}}
\usepackage{amsmath,verbatim}
\usepackage{amssymb}
\usepackage{amsfonts}

\title{
Efimov states with strong three-body losses
}

\author{F\'{e}lix Werner \thanks{\emph{Present address:} Department of Physics, University of Massachusetts, Amherst, MA 01003, USA}}
\shortauthor{F. Werner}

\institute{                    
Laboratoire Kastler Brossel, \'{E}cole Normale Sup\'{e}rieure, 
\\Universit\'{e} Pierre et Marie Curie-Paris 6, CNRS,
\\ 24 rue Lhomond, 75231 Paris Cedex 05, France
}

\pacs{21.45.-v}{Few-body systems}
\pacs{67.85.-d}{Ultracold gases}

\abstract{We
determine analytically how Efimov trimer states are modified by three-body losses
within the model of Braaten and Hammer.
We find a regime where the energies approach the positive real axis and
the decay rates vanish.
}

\begin{document}

\maketitle

%\begin{figure}
%\onefigure{epl-template.eps}
%\caption{Figure caption.}
%\label{fig.1}
%\end{figure}

\section{Introduction}

Motivated by nuclear physics,
Efimov
discovered that three particles with short-ranged 
two-body interactions of scattering length $|a|=\infty$
can support an infinite series of weakly bound trimer states~\cite{Efimov}.
In contrast to systems of nucleons or of $^4{\rm He}$ atoms,
ultracold gases of alkali metal atoms offer the possibility to
tune $a$ by using a Feshbach resonance,
and thus to go more deeply into the limit $|a|\to\infty$ 
where the Efimov effect sets in~\cite{GrimmEfimov,RevueBraaten,RevueBraaten2,EsryGreeneBurke99,EsryGammaEfi,PetrovBosons,LeeKoehler,MoraBosons}.
However the situation for alkali metal atoms is complicated by the existence
of deeply bound dimer states:
a weakly bound trimer necessarily decays into a deeply bound dimer and a free atom~\cite{EsryGammaEfi,Braaten_etats_d_efimov}.
This three-body loss process has prevented so far to produce weakly bound trimers experimentally.
The first evidence for the existence of a weakly bound trimer state is the recent observation of a peak in the three-body loss rate from an ultracold atomic Bose gas at large negative $a$~\cite{GrimmEfimov}. 
Such peaks were predicted to occur for the values of $a$ where the energy of an Efimov trimer reaches zero~\cite{EsryGreeneBurke99,Braaten_eta}.
More recently, 
a similar effect was studied experimentally and theoretically
 in a three-component Fermi gas~\cite{jochim,ohara,braaten_fermions_3_composantes,naidon_efimov}. 
The simplest description of three-body losses
is the model of
Braaten and Hammer, where 
three incoming low-energy unbound atoms are reflected back as three unbound atoms with a probability $e^{-4\eta_*}$
and recombine to a
deeply bound dimer and an atom with a probability
$1-e^{-4\eta_*}$,  $\eta_*$ being
the so-called inelasticity parameter, which depends on the short-range details of the interaction potential and of its deeply bound states~\cite{Braaten_eta,RevueBraaten,RevueBraaten2}.
This model was used to obtain
the decay rate of 
Efimov trimers  in the regime of small inelasticity parameter
$\eta_*\ll 1$~\cite{Braaten_etats_d_efimov,RevueBraaten2},
as well as
 $3$- and $4$-body scattering properties~\cite{RevueBraaten,RevueBraaten2,PetrovKokk} and the decay rate of efimovian $3$-body states in a harmonic trap~\cite{WernerTheseChap3,WernerPrepa}.
This model is expected to become exact in the zero-range limit where $|a|$ and the inverse relative momenta between atoms in the initial state are much larger than the range and effective range of the interaction potential~\cite{Braaten_eta,Braaten_etats_d_efimov,RevueBraaten,RevueBraaten2,dincao_portee_finie}.
% and than the size of deeply bound dimers

In this Letter, we determine how Efimov trimers are modified for an arbitrary inelasticity parameter $\eta_*$ by solving analytically the model of Braaten and Hammer. We find that the Efimov spectrum rotates counterclockwise in the complex plane by an angle  proportional to $\eta_*$. When $\eta_*$ reaches the critical value where this angle equals $\pi$, the discrete states disappear into the continuum. When $\eta_*$ approaches this critical value from below, the energies approach the positive real axis, so that the decay rates tend to zero.
Thus a large inelasticity parameter $\eta_*\sim1$ can paradoxically give rise to long-lived three-body states.

\begin{figure}[h]
\onefigure[width=0.6\textwidth,angle=-90]{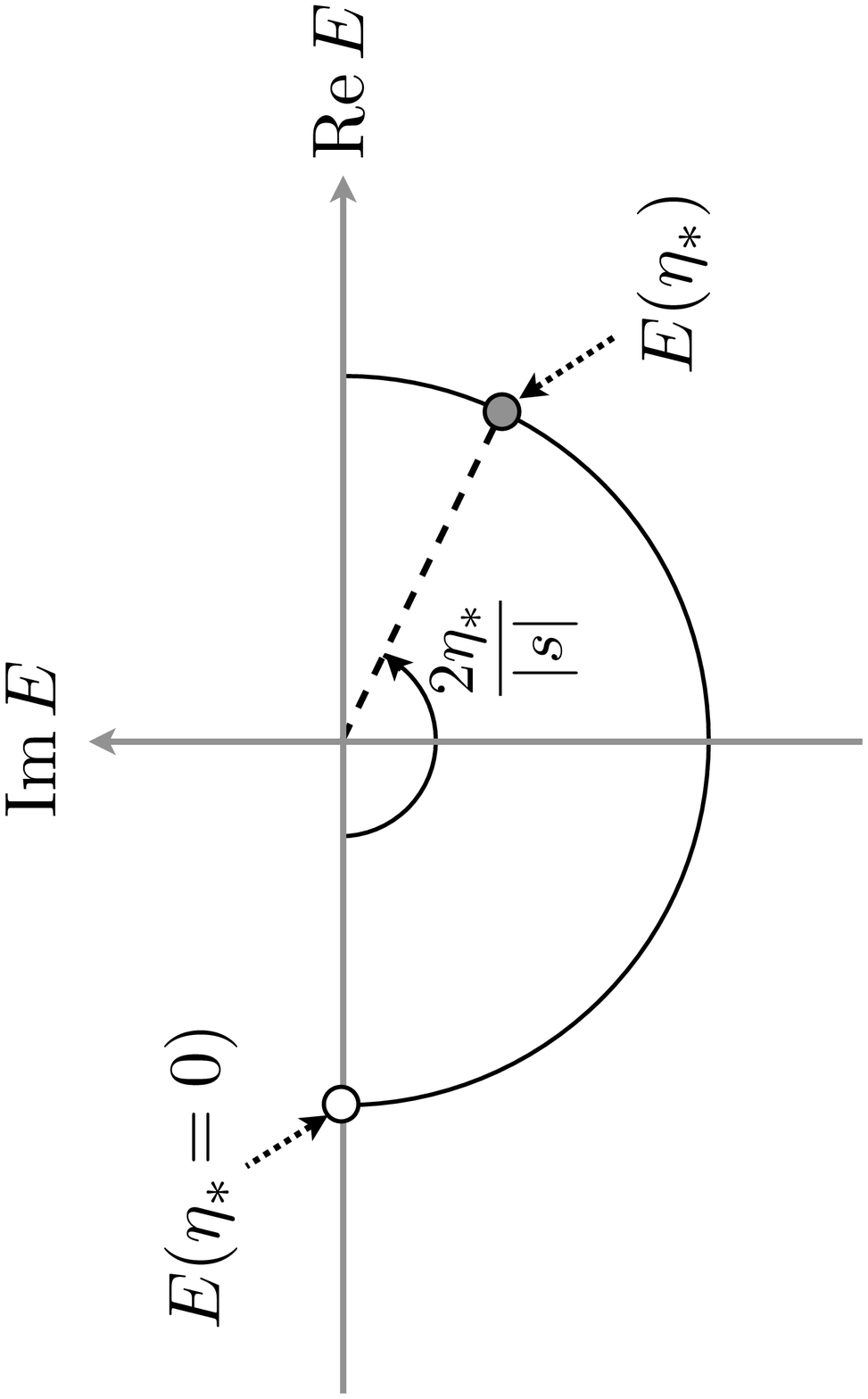}
\caption{Without three-body losses the inelasticity parameter $\eta_*=0$ and the energy of an Efimov state lies on the negative real axis.
The effect of three-body losses is to rotate this energy in the complex plane by an angle $2\eta_*/|s|$, where $|s|=1.00624\ldots$. When this angle approaches $\pi$, the decay rate
$\Gamma=-2\,{\rm Im}\, E/\hbar$ vanishes.}
\label{fig}
\end{figure}

\section{Efimov states without three-body losses}

We start by reviewing Efimov's solution of the three-body problem in the absence of three-body losses.
We restrict for simplicity to the case of three identical bosonic particles, where the wavefunction
$\psi(\mathbf{r}_1,\mathbf{r}_2,\mathbf{r}_3)$ is completely symmetric.\footnote{The case of different masses and statistics can be included without difficulty: only the value of $s$ is modified in the hyperradial problem~\cite{Efimov73}.}
In the limit where the range of the interaction potential is negligible
compared to $|a|$ and to the typical de Broglie wavelength,
the interaction can be replaced exactly by the zero-range model, see e.g.~\cite{Efimov,RevueBraaten,Albeverio3corps,AlbeverioLivre,PetrovBosons,PetrovKokk,MoraBosons,Pethick3corps,Werner3corpsPRL,WernerTheseChap3}.
An eigenstate of the zero-range model solves 
(i) the Schr\"odinger equation
\be
-\frac{\hbar^2}{2m} \sum_{i=1}^3 \Delta_{\bf r_i} \psi = E\,\psi
\label{eq:schro}
\ee
when all interparticle distances $r_{ij}$ are different from zero,
and (ii) the Bethe-Peierls boundary condition,
imposing that there exists a function $A$ such that
\be
\psi(\mathbf{r}_1,\mathbf{r}_2,\mathbf{r}_3) \underset{r_{ij}\to 0}{=} \left(\frac{1}{r_{ij}}-\frac{1}{a}\right)\,A({\bf R}_{ij},{\bf r}_k)+O(r_{ij})
\label{eq:BP}
\ee
in the limit $r_{ij}\to 0$ where particles $i$ and $j$ approach each other while the position of their center of mass ${\bf R}_{ij}$ and of the third particle ${\bf r}_k$ are fixed.
%In what follows we consider the resonant case where the scattering length diverges, $a=\infty$.
In what follows we assume that
$|a|$ is much larger than the typical distance between particles,
so that we can set $|a|=\infty$.
Equations~(\ref{eq:schro},\ref{eq:BP}) are then solved by 
Efimov's Ansatz~\cite{Efimov}
\be
\psi(\mathbf{r}_1,\mathbf{r}_2,\mathbf{r}_3)=
 F(R)
\left(1+\hat{P}_{13}+\hat{P}_{23}\right)
\frac{1}{r\rho}
\sin\left[s \arctan\left(\frac{\rho}{r}\right)\right]
\label{eq:Ansatz}
\ee
where $\hat{P}_{ij}$ exchanges particles $i$ and $j$,
$s\simeq i \cdot 1.00624$ is the only solution $s\in i
\cdot(0;+\infty)$ of the equation
$s \cos\left(s \pi/2\right) - 8/\sqrt{3}\, \sin\left(s \pi/6 \right) = 0$,
the Jacobi coordinates are $r=\|\mathbf{r}_2-\mathbf{r}_1\|$ and
$\rho=\|2\mathbf{r}_3-\mathbf{r}_1-\mathbf{r}_2\|/\sqrt{3}$,
the hyperradius is $R=\sqrt{(r^2+\rho^2)/2}$,
and the hyperradial wavefunction $F(R)$ solves the hyperradial Schr\"odinger equation
\begin{equation}
\left[-\left(\frac{d^2}{dR^2}+\frac{1}{R}\frac{d}{dR}\right)+
\frac{s^2}{R^2}
\right] F(R)= \frac{2m}{\hbar^2}\,E\,F(R).
\label{eq:schro_R}
\ee
In the limit $R\to 0$ where all three particles approach each other,
the  attractive effective potential $s^2/R^2$ diverges strongly,
and it is necessary to impose a boundary condition on the hyperradial wavefunction $F(R)$,
as first realized by Danilov~\cite{Danilov}.
This boundary condition is conveniently expressed as
\be
F(R) \underset{R\to 0}{\propto}  \left(\frac{R}{R_t}\right)^{-s}
-   \left(\frac{R}{R_t}\right)^{s}
\label{eq:Danilov}
\ee
where the three-body parameter $R_t$ depends on short-range physics
and is a parameter of the zero-range model.
The solution of Eqs.~(\ref{eq:schro_R},\ref{eq:Danilov}) is given by the
famous geometric series of Efimov trimers
\be
E_n^0=-\frac{2\hbar^2}{m R_t^{\phantom{t}2}} \, \exp\left[\frac{2}{|s|}
{\rm arg}\,\Gamma(1+s)\right]\,\exp\left(n\frac{2\pi}{|s|}\right),\ \ n\in\mathbb{Z},
\label{eq:Eefi}
\ee
with a (here unnormalized) wavefunction 
\be
F(R)=K_s(\kappa R)
\label{eq:F}
\ee
where $K$ is a Bessel  function and $\kappa$ is defined by
\be
E=-\frac{\hbar^2\kappa^2}{2m}
\label{eq:k}
\ee
with the determination
$\kappa>0$
ensuring that $F(R)$ decays exponentially for $R\to\infty$.
Efimov's spectrum~(\ref{eq:Eefi}) is unbounded from below,
in agreement with the Thomas effect~\cite{Thomas}
and with the fact that the spectrum for an interaction of finite range $b$
coincides with Efimov's spectrum only in the zero-range limit, i.e. for weakly bound trimers satisfying $|E|\ll\hbar^2/(m b^2)$~\cite{Efimov,RevueBraaten,RevueBraaten2, AmadoNoble, AmadoNoble2, Albeverio3corps,AlbeverioLattice}.

\section{Effect of three-body losses}
We now determine how Efimov's result is modified by three-body losses within the model of Braaten and Hammer. The only difference between the Braaten-Hammer model and the zero-range model is that the boundary condition (\ref{eq:Danilov}) is replaced by~\cite{RevueBraaten,RevueBraaten2}
\be
F(R) \underset{R\to 0}{\propto}  \left(\frac{R}{R_t}\right)^{-s}
-  e^{-2 \eta_*} \left(\frac{R}{R_t}\right)^{s}
\label{eq:BH}
\ee
where the inelasticity parameter $\eta_*$ and the three-body parameter $R_t$ are parameters of the Braaten-Hammer model, whose values
depend on the details of the finite-range interactions which one wishes to model.\footnote{E.g. for $^{133}{\rm Cs}$ atoms near the $-11\,{\rm G}$ Feshbach resonance,
the fit of the theoretical result of~\cite{Braaten_eta} to the experimental data of~\cite{GrimmEfimov} gives $\eta_*\simeq 0.06$~\cite{GrimmEfimov} and $R_t\simeq 30\,{\rm nm}$~\cite{GrimmEfimov,WernerTheseChap3}.}
%Again, the three-body problem with zero-range interactions is given by (\ref{eq:schro},\ref{eq:BP}), and thanks to Efimov's Ansatz (\ref{eq:Ansatz}) we reduce it to the radial
%Schr\"odinger equation (\ref{eq:schro_R}).
The physical meaning of (\ref{eq:BH}) is that
the outgoing wave $\propto R^{s}$ has an amplitude
which is smaller than the amplitude of the ingoing wave $\propto R^{-s}$
by a factor $e^{-2\eta_*}$, i.e. three ingoing atoms are reflected with a probability $e^{-4\eta_*}$
and are lost by three-body recombination with a probability $1-e^{-4\eta_*}$.

The Braaten-Hammer model is expected to become exact in the zero-range limit~\cite{Braaten_eta,Braaten_etats_d_efimov,RevueBraaten,RevueBraaten2}. 
This is supported by
numerical calculations of
the three-body loss rate from an atomic gas
 with finite-range interaction potentials, which
are in good agreement with the prediction of the Braaten-Hammer model in the zero-range regime~\cite{dincao_portee_finie}.
Moreover this can be explained using the adiabatic hyperspherical description~\cite{RevueFedorov,RevueBraaten,EsryGammaEfi,EsryPertes}: in addition to the `atomic' adiabatic channel responsible for the Efimov effect, there is one `molecular' channel associated with each deep two-body bound state;
one can thus look for decaying Gamov states with a complex energy $E$ by imposing outgoing boundary conditions in the molecular channels;
the coupling between the atomic channel and the  molecular channels
is only effective at distances on the order of the potential range,
where the wavefunction is insensitive to $E$ in the zero-range limit; and matching this short-distance wavefunction to the atomic-channel wavefunction at  hyperradii much larger than the range and much smaller than the typical atomic-channel de Broglie wavelength yields the boundary condition (\ref{eq:BH}).

In the absence of losses ($\eta_*=0$), the boundary condition (\ref{eq:BH}) reduces to Efimov's boundary condition (\ref{eq:Danilov}), and one recovers Efimov's spectrum (\ref{eq:Eefi}):
\be
E_n(\eta_*=0)=E_n^0.
\ee

In presence of losses ($\eta_*>0$), we solve the hyperradial problem (\ref{eq:schro_R},\ref{eq:BH}) with the additional boundary condition that $F(R)$ must decay quickly enough at infinity.\footnote{More precisely the normalisation condition is $\int_0^\infty dR\,R\,|F(R)|^2 < \infty$, since  the quantity $\int d{\bf r_1}\,d{\bf r_2}\,|\psi({\bf r_1},{\bf r_2},{\bf r_3})|^2 $, which does not depend on ${\bf r_3}$, has to be finite.}
Using the known properties of Bessel functions~\cite{Lebedev},
we obtain the energies
\be
E_n(\eta_*)= \exp\left(i \,\frac{2 \eta_*}{|s|}\right)\,E_n(\eta_*=0),
\label{eq:EFF}
\ee
i.~e. the spectrum is rotated in the complex plane counterclockwise around the origin by an angle $2\eta_*/|s|$.
The result (\ref{eq:EFF}) only holds if the angle $2\eta_*/|s|<\pi$, i.~e. for $\eta_*<\eta_{*c}$ with
\be
\eta_{*c} = \frac{\pi|s|}{2}=1.5806\dots,
\ee
while for $\eta_*>\eta_{*c}$ there is no normalisable solution.

The wavefunction
 is still given by Eqs.~(\ref{eq:F},\ref{eq:k}), now with the determination  
\be
{\rm Re}\, \kappa >0
\label{eq:Re>0}
\ee
of the sign of $\kappa$, which ensures that the wavefunction is normalisable since
\be
F(R)\underset{R\to\infty}{\propto}\frac{e^{-\kappa R}}{\sqrt{R}}.
\label{eq:asympt}
\ee

The decay rate
\be
\Gamma\equiv-\frac{2}{\hbar}\,{\rm Im}\, E
\ee
is given by
\be
\hbar\Gamma = 2 \sin\left( \frac{2 \eta_*}{|s|}\right)\,|E(\eta_*=0)|.
\ee
In the limit of small losses $\eta_*\ll 1$ we recover the known result~\cite{Braaten_etats_d_efimov}
\be
\hbar\Gamma\simeq \frac{4 \eta_*}{|s|}\,|E(\eta_*=0)|.
\ee
The proportionality between $\Gamma$ and the energy was first observed in numerical calculations with finite-range interactions and explained using the adiabatic hyperspherical description in~\cite{EsryGammaEfi}.

An interesting effect occurs when $\eta_*$ approaches the critical value $\eta_{*c}$ from below:
the energies approach the opposite of the energies of the Efimov states without losses
\be
E(\eta_*) \underset{\eta_*\to\eta_{*c}}{\longrightarrow}|E(\eta_*=0)|,
\label{eq:|E|}
\ee
so that the decay rates tend to zero.
Moreover, the sizes of the states diverge: since the 
imaginary part of $\kappa$
tends to a positive value
and its
real part tends to $0^+$, the behavior (\ref{eq:asympt})
of the wavefunction at large $R$ is an ingoing wave with a {\it slowly} decaying envelope.
Physically, it is not surprising that the states become increasingly delocalised before disappearing into the continuum. Since the wavefunction is normalized, this divergence of the size  implies that the probability for the three particles to be close to each other vanishes, so that $\Gamma$ vanishes.

This effect occurs within
the effective low-energy model of Braaten and Hammer.
We thus expect that it also occurs for
finite-range interactions supporting deeply bound dimers,
provided the two-body interaction potential is tuned in such a way that $\eta_*$ is slightly below $\eta_{*c}$;
this
could be checked numerically using the methods of~\cite{EsryGreeneBurke99,EsryGammaEfi,LeeKoehler,EsryPertes, dincao_portee_finie}.

Experimentally, it is rather unlikely to find a Feshbach resonance close to which
$\eta_*$ is slightly below $\eta_{*c}$. However this regime may become accessible
if interatomic interactions are tuned using an additional control parameter, e.~g.
an electric field~\cite{Kokkelmans_r_e}.

\acknowledgments

I thank E. Braaten, Y. Castin, B. Esry, H.-W. Hammer, S. Jochim, C. Mora and A. Wenz for discussions.
LKB is a {\it Unit\'{e} Mixte de Recherche} of ENS, Universit\'{e}
 Paris 6 and CNRS.
The ENS cold atoms group is a member of {\it Institut Francilien de Recherche sur les Atomes Froids}.

\bibliographystyle{eplbib}
\bibliography{felix}

\end{document}